\documentclass[aps,prl,preprint,groupedaddress]{revtex4}
\usepackage{graphicx}
\begin{document}

\title{ Intergalactic spaceflight: an uncommon way to relativistic kinematics and dynamics}
\author{Thomas Greber }
\affiliation{Department of Physics, University of Zurich, Winterthurerstrasse 190, 
CH-8057 Zurich, Switzerland} 
\author{Heinz Blatter} 
\affiliation{Institute for Atmospheric and Climate Research, ETH Zurich, 
CH-8092 Zurich, Switzerland}
\date{\today}

\begin{abstract}
In the Special Theory of Relativity space and time intervals are different  
in different frames of reference. As a consequence, the quantity ''velocity'' 
of classical mechanics splits into different quantities in Special Relativity, 
coordinate velocity, proper velocity and rapidity. The introduction and clear 
distinction of these quantities provides a basis to introduce the kinematics of 
uniform and accelerated motion in an elementary and intuitive way. Furthermore, 
rapidity links kinematics to dynamics and provides a rigorous 
way to derive Newtons Second Law in the relativistic version. Although the covariant tensorial 
notation of relativity is a powerful tool for dealing with relativistic problems, 
its mathematical difficulties may obscure the physical background of relativity 
for undergraduate students. Proper velocity and proper acceleration are the 
spatial components of the relativistic velocity and acceleration vectors, and 
thus, they provide a possibility to introduce and justify the vectorial notation 
of spacetime. The use of the three different quantities describing ''velocity'' 
is applied to discuss the problems arising in a thought experiment of a relativistic 
spaceflight.
\end{abstract}
\maketitle
\section{Introduction}

The teaching of Special Relativity is often confronted with mathematical or conceptual 
difficulties. A clear mathematical formulation using tensors is generally 
considered to be too difficult for undergraduate level. The use of simpler mathematics 
together with the verbal description of the counter-intuitive physics underlying 
Relativity also poses a great challenge for teachers and students. Verbal 
explanations trying to use pictures of classical physics in Special Relativity 
are at best confusing to the students. Several so called paradoxa arise from attempts 
to explain relativistic phenomena using a terminology coming from classical physics. 

Sometimes, one single quantity in a 
theory splits into several distinct quantities in the generalised theory. One example 
is the velocity in classical mechanics. In the widest sense, velocity or speed 
means the covered distance divided by the time needed to cover it. This is uncritical 
in classical physics, where time and distance are well defined operational concepts 
that are independent of the frame of reference, in which they are measured. In Special 
Relativity, these concepts depend on the frame of reference in which they are defined, 
if the frames are not at rest with respect to each other. This makes it necessary 
to distinguish between the different possibilities regarding the frame of reference 
in which the spatial and the temporal intervals are measured. 
This is best illustrated with a common situation of measuring the velocity of 
a rolling car. Firstly, the velocity of the car can be measured by driving past kilometer 
posts and reading the time at the moment of passing the post on synchronized watches 
mounted on the posts. Secondly, the driver can also measure the velocity by reading the 
corresponding times on a clock which is travelling with the car. Thirdly, a person with 
clock standing beside the street can measure the times on his clock at the moments, when 
the front and the rear ends of the car are passing him. The travelled distance is then taken from a measurement of the length of the car in the frame of reference of the car. A fourth 
possibility measures the velocity of the car up to an arbitrary constant
by measuring its acceleration using 
an accelerometer travelling with the car, e.g. by 
measuring a force and using Newtons Second Law, and integrates the measured acceleration 
over the time measured with a clock, also travelling with the car. 

In classical mechanics, all 
four measurements are equivalent and give the same value for the velocity. In Special 
Relativity, the first possiblility gives the coordinate velocity, which is often referred 
to as the genuine velocity. The second and third possibilities are equivalent, but are hybrid 
definitions of the speed. The temporal and spatial intervals are measured in different 
frames of reference. This speed is sometimes called celerity 
\cite{LevyLeblond&Provost1979,LevyLeblond1980}, or proper 
velocity \cite{Fraundorf1996a,Fraundorf1996b}. In addition, proper velocity is the 
spatial part of the vector of four-velocity \cite{Brehme1968}. The fourth definition 
of a speed, sometimes called rapidity \cite{LevyLeblond&Provost1979,LevyLeblond1980}, is somewhat 
distinct from the other concepts of speed in so far as it can only be determined as a change 
of speed. The need to measure an acceleration in the moving frame by means of measuring a 
force qualifies rapidity to bridge kinematics and dynamics. This seems to be not critical in 
classical mechanics, if the concept of force is accepted as an operational quantity. However, 
it can also be used to determine the relativistic version of Newtons Second Law if viewed 
from the accelerated frame of reference. 

In this paper, a way to introduce the kinematics and dynamics of Special 
Relativity is proposed by defining and consistently using the different concepts of 
speeds. This method requires the knowledge of the Lorentz transformation and the concept 
of spacetime. The kinematics of the uniform movement but also the kinematics and 
dynamics of accelerated motion can be rigorously introduced in a simple and intuitive way. 
The results are applied to an intergalactic spaceflight with constant acceleration, and the 
problems arising from relativistic effects encounterd during such a spaceflight are discussed 
both qualitatively and quantitatively.

In the following, we use
the dimensionless speeds, $\beta=v/c$ for coordinate velocity, $\omega=w/c$ for the 
proper velocity and $\rho=r/c$ for the rapidity, where $v$, $w$ and $r$ are the corresponding 
speeds measured e.g in a metric system of units for time and distance, and $c$ is 
the velocity of light in vacuum. Correspondingly, the acceleration is scaled with the speed of light, 
$\alpha=a/c$. Note that $\alpha$ has the dimension of a rate. Mass $m$ is always considered to be the invariant rest mass. 
Furthermore, we restrict our considerations to the 
2-dimensional case of spacetime with one temporal coordinate $ct$ and one spatial coordinate $x$. 
This is the simplest, non-trivial case, in which the essence of many phenomena of Special Relativity 
can be learned.

\section{The Lorentz transformation}

The hypotheses that underly the Special Theory of Relativity exclusively
concern space and time \cite{Einstein1905a,Zeeman1964,Komar1965}. 
The principle of relativity states that all inertial frames of reference are
equivalent and no absolute frame can be singled out. From these principles
an invariant velocity follows, which is in fact the speed of light.
It serves as an absolute scale and couples space and time. 
This is best seen in the Lorentz transformations that map the
coordinates $(ct_1,x_1)$ in one inertial frame of reference onto 
the coordinates $ (ct_2,x_2) $ in another inertial frame,
\
\begin{equation} \label{eq:lorentz1}
\left( \begin{array}{c} ct_2 \\ x_2 \end{array} \right) =
\left( \begin{array}{cc} \gamma & \beta \gamma \\ \beta \gamma & \gamma \end{array} \right) 
\cdot \left( \begin{array}{c} ct_1 \\ x_1 \end{array} \right),
\end{equation}
where $ \gamma = 1 / \sqrt{1-\beta^2} $ is often called relativistic factor. 
Composition of the transformations of three collinearly moving 
frames yields the known theorem of the addition of velocities,
$ \beta_{12}=(\beta_1+\beta_2)/(1+\beta_1 \beta_2) $, from which for
$ \beta_1<1 $ and $ \beta_2<1 $ results $ \beta_{12}<1 $.
The general form of a transformation corresponding to Eq. (\ref{eq:lorentz1}) is
\
\begin{equation} \label{eq:lorentz2}
\left( \begin{array}{c} ct_2 \\ x_2 \end{array} \right) =
\left( \begin{array}{cc} f(\phi) & g(\phi) \\ g(\phi) & f(\phi) \end{array} \right) 
\cdot \left( \begin{array}{c} ct_1 \\ x_1 \end{array} \right).
\end{equation}
The general parameter $\phi$ for the movement is defined by the choice of the 
functions, and can be expressed with the coordinate velocity,
\
\begin{equation} \label{eq:cond} 
\beta=g(\phi)/f(\phi).
\end{equation}
However, in most cases, 
no simple interpretation of $ \phi $ is possible. An example is given by 
$ f(\sigma)=(\sigma^2-1)/2\sigma $ and $ g(\sigma)=(\sigma^2+1)/2\sigma $,
where the relation of $ \sigma $ with the coordinate velocity is
$\beta=(\sigma^2-1)/(\sigma^2+1) $. 
Composition of transformations yields that $ \sigma $
is multiplicative, $ \sigma_{12}=\sigma_1 \sigma_2 $. Furthermore,
$\sigma = \sqrt{(1+\beta)/(1-\beta)}$ is the Doppler shift.
Another example uses the hyperbolic functions \cite{Brehme1968},
\
\begin{equation} \label{eq:lorentz3}
\left( \begin{array}{c} ct_2 \\ x_2 \end{array} \right) =
\left( \begin{array}{cc} \cosh (\rho) & \sinh (\rho) \\ \sinh (\rho) & \cosh (\rho) \end{array} \right) \cdot \left( \begin{array}{c} ct_1 \\ x_1 \end{array} \right).
\end{equation}
From Eqs. (\ref{eq:lorentz1}) and (\ref{eq:lorentz3}) follows $ \beta=\tanh (\rho) $. 
Composition of transformations yields the additivity of the parameter, 
$ \rho_{12}=\rho_1+\rho_2 $, and thus the relation $ \rho = \ln (\sigma) $ to 
the multiplicative parameter $ \sigma $. The additivity of $ \rho $ was already 
mentioned by \cite{Pars1921}.

\section{Kinematics}

\subsection{Speeds}

Motion is generally described with a velocity, i.e. the differential limit
\
\begin{equation} \label{eq:speed1}
\beta=\lim_{\Delta t \rightarrow 0}\,\frac{\Delta x}{c\,\Delta t},
\end{equation}
where $\Delta x$ is a change in position within the time step $\Delta t$,
and $c$ is the speed of light.
In Galilean spacetime, $\Delta t$ and $\Delta x$ are the same in all 
frames of reference. In relativistic spacetime this is not the case and a
careful distinction of different operational definitions of speed has
to be made. It is important to distinguish whether the two quantities 
$\Delta x$ and $\Delta t$ are measured in the same inertial frame or not.
As mentioned above, the latter case is e.g. applied by car drivers for the calibration 
of the tachometer. For the proper velocity the time step $\Delta \tau$ 
is measured on the clock travelling with the car and the  change in position 
(kilometer posts) $\Delta x$ is measured in the rest frame of the street,
\
\begin{equation} \label{eq:speed2}
\omega = \lim_{\Delta \tau \rightarrow 0}\,\frac{{\Delta x}}{{c\,\Delta \tau}}, 
\end{equation}
The reciprocal measurement yields the same result: an observer outside
the car measures the time $\Delta t$ on her clock between the passages of the 
front and rear ends of the car, however, using the proper length $\Delta l$ of
the car as specified by the manufacturer,
\
\begin{equation} \label{eq:speed2a}
\omega = \lim_{\Delta t \rightarrow 0}\,\frac{\Delta l}{c\,\Delta t}.
\end{equation}
Since $\Delta t=\gamma\,\Delta\tau$ and $\Delta l=\gamma\Delta x$, the
proper velocity is related to the coordinate velocity by $\omega=\gamma \beta$,
and $\gamma = \sqrt{1+\omega^2}$. 
The proper velocity $\omega$ is an unbound quantity. 

Another possibility to quantify motion is using quantities measured in the 
accelerated vehicle alone. The rapidity of a spacecraft is 
defined as the integral of the correspondingly recorded acceleration $\alpha$ 
with respect to proper time $\tau$ \cite{LevyLeblond&Provost1979,LevyLeblond1980},
\
\begin{equation} \label{eq:speed3}
\rho =  \int \alpha(\tau)\, {\rm d}{\tau} + \rho_0,
\end{equation}
where the rapidity $\rho$ is defined up to an integration constant $\rho_0$. This measurement is not critical in classical physics. In relativistic physics, the observability of this acceleration is not obvious and needs further analysis.

\subsection{Acceleration}

In classical physics, acceleration is defined by
\
\begin{equation} \label{eq:acceler1}
\alpha = \lim_{\Delta t \rightarrow 0}\,\frac{\Delta v}{c\,\Delta t},
\end{equation}
In relativistic physics, several possibilities exist whether for $\Delta v /c$ the coordinate velocity $\Delta \beta$, the proper velocity $\Delta \omega$ or the rapidity $\Delta \rho$, and whether the coordinate time $\Delta t$ or the proper time $\Delta \tau$ is chosen.

To introduce different definitions of acceleration, a spacecraft is considered, that changes its velocity in a given time interval. Two inertial frames of reference, I$_1$ and I$_2$, are defined, co-moving with the spacecraft at the beginning and at the end of the time interval, respectively. The coordinate velocities of I$_1$ and I$_2$ with respect to a chosen inertial frame of reference are $\beta_1$ and $\beta_2$, and the corresponding velocity increment is $\Delta \beta = \beta_2 - \beta_1$. The relative velocity $\Delta \beta_{12}$ between I$_1$ and I$_2$ is given by the Einsteinian velocity combination law,
\
\begin{equation} \label{eq:addition}
\Delta \beta_{12} = \frac{\beta_2-\beta_1}{1 -\beta_1\beta_2} 
= \frac{(\beta_1+\Delta \beta)-\beta_1}{1 -(\beta_1+\Delta \beta)\beta_1}
\end{equation}
In the differential limit for vanishing length of the time interval, 
\
\begin{equation} \label{eq:speed4}
{\rm d}\beta_{12} =  \gamma^2\,{\rm d}\beta.
\end{equation}
The acceleration $\alpha$ defined by the velocity increment $\Delta \beta_{12}$ and the proper time interval $\Delta \tau$ are proper quantities determined in the frame of the spacecraft alone  \cite{LevyLeblond1980},
\begin{equation} \label{eq:acceler2}
\alpha = \frac{{\rm d}\beta_{12}}{{\rm d}\tau} = 
\frac{1}{{\rm d}\tau} ( \gamma^2\,{\rm d}\beta).
\end{equation}
and by integration,
\
\begin{equation} \label{eq:speed5}
\rho = \int \alpha {{\rm d}\tau'} = \int \gamma^2\,{\rm d}\beta' = 
{\rm arctanh}\, (\beta) + \rho_0.
\end{equation}
the rapidity $\rho$ is defined up to an integration constant $\rho_0$.
Similar to proper velocity, rapidity is an unbound quantity. Comparing Eq. (\ref{eq:speed5}) with Eqs. (\ref{eq:cond}) and (\ref{eq:lorentz3}), with an appropriate choice of the frame of reference, such that $\rho_0=0$ we find 
\
\begin{equation} \label{eq:rap1}
\gamma = \cosh (\rho) \quad {\rm and} \quad
\omega = \gamma \beta = \sinh (\rho). 
\end{equation}

With Eq. (\ref{eq:rap1}), the proper acceleration $\alpha$ is
\
\begin{equation} \label{eq:acc1}
\alpha = \frac{{\rm d}\rho}{{\rm d}\tau}=
\frac{{\rm d}}{{\rm d}\tau} {\rm arcsinh}\,(\omega)=
\frac{1}{\sqrt{1+\omega^2}}\ \frac{{\rm d} \omega}{{\rm d} \tau}=
\frac{{\rm d} \omega}{{\rm d}t} =
\gamma^3\,\frac{{\rm d} \beta}{{\rm d}t},
\end{equation}
where ${{\rm d} \beta}/{{\rm d}t}$ is the coordinate acceleration.
This is an interesting and useful result. The acceleration $\alpha$ is not only the
derivative of rapidity $ \rho $ with respect to the proper time $ \tau $, but 
is equal to the derivative of the proper velocity $ \omega $ with respect to coordinate 
time $t$ in an inertial reference frame. This corresponds to the definition of proper acceleration, which corresponds to the spatial part of the 
relativistic acceleration vector.

Equation (\ref{eq:acc1}) suggests to define the uniform accleration 
as uniform proper acceleration, ${\rm d} \omega/{\rm d}t = const$, rather than a 
constant coordinate acceleration, ${\rm d} \beta/{\rm d}t$. If the velocity 
of the spacecraft is $ \beta_0 = \omega_0 = \rho_0 = 0 $ at time $ t_0 = \tau_0 = 0 $, 
the rapidity after uniform acceleration $\alpha$ at proper time $ \tau $ is 
$ \rho = \alpha \tau $, and with Eq. (\ref{eq:speed2}), 
the distance travelled in an inertial frame of reference is
\
\begin{equation} \label{eq:acc2a}
x = \int_0^x\,{{\rm d}x'}=c \int_0^\tau\, \omega(\tau')\,{{\rm d}\tau'}=
c \int_0^\tau\, \sinh(\alpha \tau')\,{{\rm d}\tau'}=
\frac{c}{\alpha}\, \left[ \cosh(\alpha \tau)-1 \right].
\end{equation}

Conversely, the proper time needed to travel the distance $ x $ in the reference system is

\begin{equation} \label{eq:acc3}
\tau = \frac{1}{\alpha}\, {\rm arccosh}\left( 1+\frac{\alpha x}{c}\right).
\end{equation}
Integration of Eq. (\ref{eq:acc1}) yields the needed coordinate time to reach the 
proper velocity $ \omega $,
\

\begin{equation} \label{eq:acc4}
t=\frac{1}{\alpha}\,\int_0^{\omega}\,{{\rm d} \omega'} =
\frac{\omega}{\alpha}=\frac{1}{\alpha}\, \sinh(\alpha \tau).
\end{equation}

\begin{figure} [hbt] 
\vspace{-1.5cm}
\hspace{1.5cm}
\includegraphics[width=140mm]{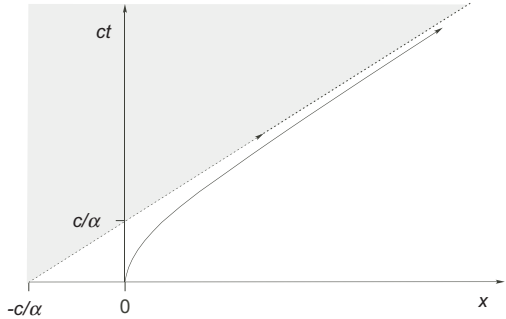}
\caption{World line of an uniformly accelerating spacecraft (solid line) and of a photon (dashed line), both starting at the same time on the x-axis. The shaded area lies behind a horizon, from where no information can reach the spacecraft.}
\label{f1}
\end{figure}

Equations (\ref{eq:acc2a}) and (\ref{eq:acc4}) constitute a parameter equation for the 
world line of an uniformly accelerated spacecraft. 
Elimination of the parameter $ \alpha $ yields the coordinate equation,
\
\begin{equation} \label{eq:acc5}
\left( x+\frac{c}{\alpha}\right)^2 - (ct)^2 = \left( \frac{c}{\alpha}\right)^2,
\end{equation}
which is the equation of a hyperbola in a Minkowski diagram. The asymptotes of the 
hyperbola have inclinations $\pm 1$ and are parallel to the light cone. Figure \ref{f1} illustrates  
an interesting consequence: a spacecraft starting at $ x=0 $ moving with constant 
proper acceleration $ \alpha $ in the direction of the positive $x$-axis outruns a photon 
starting simultaneously at $ x \le -c/\alpha $ in the same direction (Misner et al., 1973).

\section{Relativistic dynamics}

The spacetime of Einsteinian Relativity makes a revision of Newtonian 
mechanics necessary. Newtons Second Law involves more than one frame of
reference and this turns out to have consequences on the dynamics.
Newtons Second Law $F={\rm d}(m\,v)/{{\rm d}t}$, where $F$ is the 
force acting during the time increment ${\rm d}t$ on the mass
$m$ that moves with the velocity $v$ is rewritten as
\
\begin{equation} \label{eq:feynman}
F= \frac{{\rm d}(\gamma m\,v)}{{\rm d}t}.
\end{equation}
In relativistic spacetime it has to be specified that the time
increment ${\rm d}t$ is measured in the frame of reference in which the
mass is moving with velocity $v$.
Equation. (\ref{eq:feynman}) is the correct equation and has the property that all
physical quantities are defined in the same frame of reference.

In this paper a derivation of relativistic dynamics is suggested
in which Newtons Second Law is written in quantities described in the rest
frame of the accelerated mass. These quantities are the invariant
mass, the proper time and the rapidity. 
For simplicity, the derivation is restricted to linear motion and collinear
acceleration.

Rapidity needs a concept of inertia for the measurement of proper acceleration,
such that an accelerometer measures acceleration independent of the actual
speed with respect to any inertial frame of reference. If this were not the
case, then one specific inertial frame could be singled out to serve as
an absolute frame of reference, and the principle of relativity would be
violated.  A measurement of a weight in a laboratory or in a spacecraft
flying in space is a static measurement with respect to the spacecraft, 
and Newtons Second Law applies exactly. An astronaut
feeling a constant weight thus correctly interpretes the motion as uniformly
accelerated. With a gyroscope, it is possible to distinguish between a linearly 
accelerated or purely circular motion, or any combination of both.

In the simplest case of linear acceleration, the astronaut interpretes the
proper acceleration as a change of the rapidity per proper time interval.
Furthermore, the constant weight of the astronaut is correctly interpreted 
as the constant force needed to accelerate the invariant rest mass with
the constant proper acceleration, independent of the rapidity with
respect to any desired frame of reference.
In the case of non-uniform linear acceleration, the momentary weight $F$ 
can be interpreted as the constant force needed to accelerate
the mass $m$ of the astronaut with the momentary proper acceleration
$\alpha$, again independent of the momentary rapidity,
\
\begin{equation} \label{eq:force1}
F = m\,c\,\alpha = m\,c \, \frac{{\rm d}\rho}{{\rm d}\tau} = m\,c\,\frac{{\rm d} \omega}{{\rm d}t} = m\,c \, \gamma^3\ \frac{{\rm d}\beta}{{\rm d}t} = 
\frac{{\rm d}(\gamma m\,v)}{{\rm d}t},
\end{equation}
which is the relativistic ''Second Law'', Eq. (\ref{eq:feynman}), written in terms of velocity and time
as measured in an inertial frame of reference moving with velocity $v=c\,\beta$ 
relative to the astronaut with mass $m$.
This makes it particularly transparent that the factor $\gamma^3$ in the
relativistic relation between force and the acceleration stems from the
transformation of space and time, and has no physical relation to the mass.

Using the proper quantities, the notation of classical kinematics can be recovered 
for relativistic kinematics, Eqs. (\ref{eq:speed2a}) and (\ref{eq:acc1}), such that 
the proper velocity is the derivative of proper length with respect to coordinate 
time and proper acceleration is the derivative of proper velocity with respect to 
coordinate time. This pattern can be extended to dynamics be defining force $F$ as 
the derivative of linear momentum $p$ with respect to coordinate time. From 
Eq. (\ref{eq:force1}) we get
\
\begin{equation} \label{eq:momentum1}
F = m\,c\,\frac{{\rm d} \omega}{{\rm d}t} \equiv \frac{{\rm d} p}{{\rm d}t},
\end{equation}
and consequently, the linear momentum is $p = m\,c\,\omega$. The kinetic energy of 
a relativistic particle is best derived by computing 
the work applied to the particle to accelerate it to a velocity 
$\beta$ \cite{Tipler1999},
\
\begin{equation} \label{eq:energy2}
E_\mathrm{kin} = \int^d_0 F \ \mathrm{d}x
\end{equation}
where the particle travels a distance $d$ in the system where $E_{\mathrm{kin}}$ is measured. Applying the relativistic Second Law 
(Eq. \ref{eq:force1}), we obtain 
\
\begin{equation} \label{eq:energy3}
E_\mathrm{kin} 
= m\,c\,\int^d_0 \gamma'^3 \frac{\mathrm{d}\beta'}{\mathrm{d}t}\ \mathrm{d}x
= m\,c^2\,\int^{\beta'}_0 \beta'\,\gamma'^3 \ \mathrm{d}\beta' =
m\,c^2\,\left( \gamma - 1  \right),
\end{equation}
where the dashed variables $\beta'$ and $\gamma'$ denote the integrations variables, and
the kinetic energy $E_\mathrm{kin}$ is a function of the final $\gamma$.

\section{Space Travel}

To reach interstellar distances within a reasonable time, high velocities, and thus, 
continuous acceleration during travel time are required. A flight to a star must 
include two stages: acceleration to high velocities and breaking down the velocity 
to the velocity of the star. At a constant proper acceleration of 9.8 m/s$^2$, corresponding 
to about 1 Lightyear/year$^2$, a flight to $\alpha$-Centauri in 4.3 Lightyears 
distance will last about 3.63 proper years. In the middle of the trip, the rapidity 
of the spacecraft would be $ \rho=1.4 $, the proper velocity $ \omega=1.9$ and the coordinate 
velocity relative to Earth $ \beta=0.88 $. A flight to $\alpha$-Centauri and back 
to Earth would last at least 7.26 proper years, whereas on Earth 14.5 years would pass 
between take off and return of the spacecraft. The corresponding times to travel to the 
Andromada Galaxy and back is 58 proper years, and the corresponding speeds in the 
middle between Earth and the Andromada Galaxy are $ \rho=14.5 $, 
$ \omega=991380 $ and $ \beta=0.9999999999995=1-5 \cdot 10^{-13} $. On Earth, about 
8 Million years elapsed during the corresponding trip to the Andromeda galaxy in a 
distance of 2 Million Lightyears from Earth, and back. 
At this constant proper acceleration, the covered distances become large at a large 
rate. In 20 proper years, the astronaut covers a distance of 250 Million Lightyears, 
36 Billion Lightyears after 25 proper years. It is in principle possible to travel 
to the most distant galaxies within a human life time, as is outlined quite 
realistically in the science fiction novel ''Tau Zero'' by \cite{Anderson1970}, 
if such a spacecraft would be available.

However, the realization of this cosmic travel plan requires the solution of some 
serious problems, of which most are based on relativistic effects. The first problem 
concerns the necessary specification for the spacecraft that is able to accelerate 
over a long time in outer space. There are two different possibilities: the first is 
a rocket that carries all necessary fuel and energy from the beginning, and the second 
is the ramjet \cite{Bussard1960,Semay&Silvestre2005} that collects interstellar 
matter that can be used for fuel and energy supply. 

Many problems are caused by the existing radiation in space and the interstellar or 
intergalactic matter. Although the spacecraft travels through almost empty space, 
problems are caused by the extreme Doppler shift of the electromagnetic radiation at 
the high relativistic speeds, and even more, by the existing matter mostly in the 
form of hydrogen and helium atoms and dust particles.

\subsection{Interstellar matter}

The density of interstellar matter is estimated to be about one hydrogen atom per 
cubic centimeter within galaxies. Between the galaxies, it is about 6 orders of 
magnitude smaller with about 1 hydrogen atom per cubic meter. Within dense nebulae, 
particles made of ice and carbon of 10$^{-18}$ kg may occur \cite{Spitzer1978}, 
however, they contribute only to about half a percent of the total mass. 

We assume that the spacecraft does not deflect the interstellar matter by using e.g. 
a magnetic field. In this case, the particles and atoms are stopped and collected by 
the ship. To calculate the particle flux $ \dot N $, the proper velocity 
$ \omega $ is the adequate quantity. The number 
of collected particles depends on the volume $ {\rm d}V $ of space covered by the spacecraft 
as measured in the reference frame of Earth, however, seen from the spacecraft, the flux must
be measured in units of proper time $ {\rm d}\tau $. The volume $ \mathrm{d}V $covered in a 
proper time interval $ \mathrm{d}\tau $ by a spacecraft with a cross sectional area $A$ at
a proper velocity $\omega$ is
\
\begin{equation} \label{eq:mass2}
{\rm d}V = A \omega c\, {\rm d} \tau,
\end{equation}
and the corresponding particle flux 
\
\begin{equation} \label{eq:mass3}
\dot N_r = A \omega c n,
\end{equation}
where $ n $ is the density of particles. 
With the assumption that the majority of particles are hydrogen atoms with a rest mass 
of $ \mu_0 = 1.7 \cdot 10^{-27} $ kg, and with a particle density in intergalactic 
space of $ n_{\rm intergalactic} = 1 $ m$^{-3}$, the particle flux on the spacecraft 
at a velocity of  $ \omega = 10^6 $ becomes $ \dot N =3 \cdot 10^{14} $ s$^{-1}$m$^{-2}$, 
and the corresponding mass flux is $ 5 \cdot 10^{-13} $ kg\,s$^{-1}$m$^{-2}$. 

The momentum $ p_0 $ of a particle with mass $ \mu_0 $ is
\
\begin{equation} \label{eq:mass4}
p_0 = \mu_0 \omega c,
\end{equation}
and thus, the pressure $P$ on the front of the ship exerted by this particle flux increases 
with the square of the proper velocity,
\
\begin{equation} \label{eq:mass6}
P = \frac{\dot N_0 p_0}{A} = \mu_0 n \omega^2 c^2,
\end{equation}
corresponding to 150 Pa for the above situation. The kinetic energy that each particle 
deposits in the ship is 
\
\begin{equation} \label{eq:energy}
E_{\rm kin} = \mu_0 c^2 (\gamma-1).
\end{equation}
For a proton this corresponds 
to about 1 TeV at $ \omega = 10^6 $. This energy can be reached in todays most powerful 
accelerators, such as Tevatron at Fermilab. The corresponding energy flux at the front 
of the ship is then $4.5 \cdot 10^{10}$ Wm$^{-2}$, corresponding to a black body 
radiation with a temperature of nearly 30'000 K.

Equation (\ref{eq:energy}) also indicates a problem of the Bussard-ramjet. 
The kinetic energy of
the collected particles eventually exceeds their rest energy,
\
\begin{equation} \label{eq:energy1}
E_{\rm kin} = \mu_0 c^2 (\gamma-1) > \mu_0 c^2.
\end{equation}
Therefore, even if all of the rest energy of collected particles could be applied to accelerate 
the jet to propulse the spacecraft, its speed could never exceed $ \gamma=2 $. Realistically, 
a nuclear fusion reactor gains about 1 per mille of the rest energy. This limits 
the speed of such a ramjet to about 3\% of the speed of light. A Bussard-ramjet 
could only operate at higher velocities if all the collected particles are funneled through the engines without stopping them. They would then have to be accelerated 
by using an energy source that has to be carried with the ship from the 
beginning of the journey.

\subsection{Electromagnetic radiation}

The spacecraft flies through a field of electromagnetic radiation. The sources of this 
radiation are the stars and the cosmic microwave background, and its spectrum ranges 
from radio waves to gamma rays and beyond. The energy of the cosmic microwave background 
dominates the background radiation in deep space far away from stars
\cite{Sandage&Kron1993}. At relativistic 
speeds the radiation field is strongly changed due to Doppler shift and aberration 
\cite{Komar1965,Blatter&Greber1988}. The 
astronaut registers a radiation field of high intensity and shifted to high frequencies 
around the apex, i.e. in the direction of the journey, and very little radiation 
from the rest of the celestial sphere.

The Doppler shift in the direction of the apex is $ f' = f\,(\gamma+\omega)$,
where $f$ and $f'$ are the frequencies of the radiation in the rest frame of the radiation 
source and seen from the spacecraft, respectively. At $\omega\approx\gamma=10^6$, as in the middle of 
the trip to the Andromada galaxy, the cosmic background radiation of a wavelength of 1 mm 
is shifted to soft X-rays with a wavelength of 5 {\AA}ngstr\"om. The pressure of this radiation 
on the front of the spacecraft is $7 \cdot 10^{-8}$ Pa \cite{Greber&Blatter1990}, 
which is negligibly small compared with the pressure exerted by the intergalactic matter. 
However, When heading towards a star, the Doppler shifted stellar radiation would become extremely hard and living organisms must be shielded from this radiation.

\section{Discussion}

The consequent distiction and application of the three types of speeds in 
teaching Special Relativity allows us to explain the relativistic kinematics and 
dynamics in an intuitive way. The description of motion, acceleration and linear 
momentum can be recovered in the classical way by replacing the coordinate velocity 
with the proper velocity. This fact is not novel since in the vectorial notation, 
the spatial parts of velocity and acceleration vectors correspond to proper 
velocity and proper acceleration. However, this is often obscured by taking the 
relativistic factor $\gamma$ out of the vector components, thus writing the 
vectors as the relativistic factor times a vector with the components of the 
coordinate velocity.

We used the facts that e.g. an astronaut in a spacecraft could not perceive a uniform 
motion of his spacecraft without looking out of a window, and he would perceive a constant 
proper acceleration by feeling his own constant weight independent, however, of his velocity 
relative to any external frame of reference. These situations can be mapped to 
everyday experiences in traveling with trains or airplanes, and thus, can be based 
on the classical concepts of motion and inertia. The application of the relativistic 
kinematics and dynamics to steady accelerated spaceflight may make the topic 
more appealing to students than misleading paradoxa and difficult concepts such 
as length contraction and time dilatation. 

The restriction to the two-dimensional spacetime, one temporal and one spatial 
coordinate, limits the application of this method to collinear motion, acceleration 
and forces. On the other hand, the results offer the basis for the vectorial 
notation in two dimensions, which then can readily be extended to the 1+3 dimensional 
general case. The topic may then be extendend to what the astronaut really observes
if he looks out of the window of his spacecraft, not length contraction, but aberration 
and Doppler shift of electromagnetic waves, and it only makes sense to compare the 
different times that passed in the spacecraft and on Earth when he returns to Earth. 

\section*{Acknowledgements}

The authors thank Joachim Stadel, who reviewed an earlier version of the paper and helped to improve it substantially.


\end{document}